\documentclass[12pt]{article}
\usepackage{amsthm,amsmath,amssymb}
\usepackage{url}
\usepackage{srcltx} 
\usepackage{enumerate}
\usepackage[utf8]{inputenc}
\usepackage{mathrsfs}
\usepackage[ruled,vlined]{algorithm2e}
\usepackage{xspace}
\usepackage{framed}
\usepackage{enumerate}
\usepackage{xcolor}
\usepackage{xspace}
\usepackage[colorlinks=true,citecolor=black,linkcolor=black,urlcolor=blue]{hyperref}

\newcommand{\arxiv}[1]{\href{http://arxiv.org/abs/#1}{\texttt{arXiv:#1}}}

\def\abs#1{\left| #1 \right|}
\def\paren#1{\left( #1 \right)}
\def\acc#1{\left\{ #1 \right\}}

\def\cro#1{\left[ #1 \right]}

\theoremstyle{plain}
\newtheorem{theorem}{Theorem}

\theoremstyle{definition}

\theoremstyle{remark}



\begin{document}

\title{Doubled patterns are $3$-avoidable}

\author{Pascal Ochem\\
\small LIRMM, Universit\'e de Montpellier, CNRS\\[-0.8ex]
\small Montpellier, France \\[-0.8ex] 
\small\tt ochem@lirmm.fr\\
}
\maketitle

\begin{abstract}
  In combinatorics on words, a word $w$ over an alphabet $\Sigma$ is
  said to avoid a pattern $p$ over an alphabet $\Delta$ if there is no
  factor $f$ of $w$ such that $f=h(p)$ where $h:
  \Delta^*\to\Sigma^*$ is a non-erasing morphism. A pattern $p$ is
  said to be $k$-avoidable if there exists an infinite word over a
  $k$-letter alphabet that avoids $p$. A pattern is said to be doubled
  if no variable occurs only once. Doubled patterns with at most 3 variables and
  patterns with at least 6 variables are $3$-avoidable. We show that doubled
  patterns with 4 and 5 variables are also $3$-avoidable.

  %\bigskip\noindent \textbf{Keywords:} Word; Pattern avoidance.
\end{abstract}

%%%%%%%%%%%%%%%%%%%%%
\section{Introduction}\label{sec:intro}
%%%%%%%%%%%%%%%%%%%%%

A pattern $p$ is a non-empty word over an alphabet
$\Delta=\{A,B,C,\dots\}$ of capital letters called \emph{variables}.
An \emph{occurrence} of $p$ in a word $w$ is a non-erasing morphism $h:\Delta^*\to\Sigma^*$
such that $h(p)$ is a factor of $w$.
The avoidability index $\lambda(p)$ of a pattern $p$ is the size of the
smallest alphabet $\Sigma$ such that there exists an infinite word $w$
over $\Sigma$ containing no occurrence of $p$.
Bean, Ehrenfeucht, and McNulty~\cite{BEM79} and Zimin~\cite{Zimin}
characterized unavoidable patterns, i.e., such that $\lambda(p)=\infty$.
We say that a pattern $p$ is $t$-avoidable if $\lambda(p)\le t$.
For more informations on pattern avoidability, we refer to Chapter 3 of Lothaire's book~\cite{Lothaire:2002}.

It follows from their characterization that every unavoidable pattern contains a variable that occurs once.
Equivalently, every doubled pattern is avoidable.
Our result is that :
\begin{theorem}
\label{2tok}
Every doubled pattern is $3$-avoidable.
\end{theorem}

Let $v(p)$ be the number of distinct variables of the pattern~$p$.
For $v(p)\le3$, Cassaigne~\cite{Cassaigne:1994} began and I~\cite{Ochem:2004}
finished the determination of the avoidability index of every pattern with at most 3 variables.
It implies in particular that every avoidable pattern with at most 3 variables is 3-avoidable.
Moreover, Bell and Goh~\cite{BellGoh:2007} obtained that every doubled pattern $p$
such that $v(p)\ge6$ is 3-avoidable.

Therefore, as noticed in the conclusion of~\cite{OchemPinlou}, there remains to prove Theorem~\ref{2tok} for every pattern $p$
such that $4\le v(p)\le5$. In this paper, we use both constructions of infinite words
and a non-constructive method to settle the cases $4\le v(p)\le5$.

Recently, Blanchet-Sadri and Woodhouse~\cite{Blanchet:2013} and Ochem and Pinlou~\cite{OchemPinlou}
independently obtained the following.

\begin{theorem}[\cite{Blanchet:2013,OchemPinlou}]
\label{BGR}
Let $p$ be a pattern.
\begin{enumerate}[$\quad$(a)]
\item If $p$ has length at least $3\times2^{v(p)-1}$ then $\lambda(p)\le2$.\label{a}
\item If $p$ has length at least $2^{v(p)}$ then $\lambda(p)\le3$.\label{b}
\end{enumerate}
\end{theorem}
As noticed in these papers, if $p$ has length at least $2^{v(p)}$ then $p$ contains a doubled
pattern as a factor. Thus, Theorem~\ref{2tok} implies Theorem~\ref{BGR}.(\ref{b}).

%%%%%%%%%%%%%%%%%%%%%
\section{Extending the power series method}\label{sec:method}
%%%%%%%%%%%%%%%%%%%%%

In this section, we borrow an idea from the entropy compression method to
extend the power series method as used by Bell and Goh~\cite{BellGoh:2007},
Rampersad~\cite{Rampersad:2011}, and Blanchet-Sadri and Woodhouse~\cite{Blanchet:2013}.

Let us describe the method.
Let $L\subset\Sigma_m^*$ be a factorial language defined by a set $F$ of forbidden factors
of length at least 2.
We denote the factor complexity of $L$ by $n_i=L\cap\Sigma_m^i$.
We define $L'$ as the set of words $w$ such that $w$ is not in $L$ and the prefix
of length $|w|-1$ of $w$ is in $L$.
For every forbidden factor $f\in F$, we choose a number $1\le s_f\le|f|$.
Then, for every $i\ge 1$, we define an integer $a_i$ such that
$$a_i\ge\max_{u\in L}\abs{\acc{v\in\Sigma_m^i\ |\ uv\in L',\ uv=bf,\ f\in F,\ s_f=i}}.$$
We consider the formal power series $P(x)=1-mx+\sum_{i\ge1}a_ix^i$.
If $P(x)$ has a positive real root $x_0$, then $n_i\ge x_0^{-i}$ for every $i\ge0$.
%Notice that $P(x)$ has at most two positive real roots.
%If $P(x)$ has two positive real roots $x_0$ and $x_1$ (with $x_0<x_1$), then $L\cap\Sigma_m^i\ge x_0^{-i}$ for every $i\ge0$.
%If $P(x)$ has a positive real double root $x_0$, then $L\cap\Sigma_m^i\ge(i+1)x_0^{-i}$ for every $i\ge0$.

Let us rewrite that $P(x_0)=1-mx_0+\sum_{i\ge1}a_ix_0^i=0$ as 
\begin{equation}\label{p}
 m-\sum_{i\ge1}a_ix_0^{i-1}=x_0^{-1}
\end{equation}
Since $n_0=1$, we will prove by induction that $\frac{n_i}{n_{i-1}}\ge x_0^{-1}$ in order to obtain that $n_i\ge x_0^{-i}$ for every $i\ge0$.
By using (\ref{p}), we obtain the base case: $\frac{n_1}{n_0}=n_1=m\ge x_0^{-1}$.
Now, for every length $i\ge1$, there are:
\begin{itemize}
 \item $m^i$ words in $\Sigma_m^i$,
 \item $n_i$ words in $L$,
 \item at most $\sum_{1\le j\le i}n_{i-j}a_j$ words in $L'$,
 \item $m(m^{i-1}-n_{i-1})$ words in $\Sigma_m^i\setminus\acc{L\cup L'}$.
\end{itemize}
This gives $n_i+\sum_{1\le j\le i}n_ja_{i-j}+m(m^{i-1}-n_{i-1})\ge m^i$, that is, $n_i\ge mn_{i-1}-\sum_{1\le j\le i}n_{i-j}a_j$.
%$\frac{n_i}{n_{i-1}}\ge m-\sum_{1\le j\le i}a_j\frac{n_{i-j}}{n_{i-1}}\ge m-\sum_{1\le j\le i}a_jx_0^{j-1}\ge m-\sum_{j\ge1}a_jx_0^{j-1}=x_0^{-1}$.

$$\begin{array}{llll}
\frac{n_i}{n_{i-1}} & \ge & m-\sum_{1\le j\le i}a_j\frac{n_{i-j}}{n_{i-1}} &\\
 & \ge & m-\sum_{1\le j\le i}a_jx_0^{j-1} & \text{ By induction }\\
 & \ge & m-\sum_{j\ge1}a_jx_0^{j-1} & \\
 & = & x_0^{-1} & \text{ By (\ref{p})}
\end{array}$$

% If $x_0$ is a double positive real root of $P$, then ?????.
% To prove this, we first observe that $P'(x_0)=0$, which gives
% \begin{equation}\label{p'}
% m=\sum_{i\ge1}ia_ix_0^{i-1}.
% \end{equation}
% By combining (\ref{p}) and (\ref{p'}), we obtain
% \begin{equation}\label{pp'}
% \sum_{i\ge1}(i-1)a_ix_0^{i-1}=x_0^{-1}. 
% \end{equation}
% We show by induction that $\frac{n_i}{n_{i-1}}\ge\frac{i}{i-1}x_0^{-1}$ for every $n\ge2$.
% Recall that $n_0=1$, $n_1=m$, and $n_2\ge m^2-a_1m-a_2$.
% We have that
% $$\begin{array}{llll}
% \frac{n_i}{n_{i-1}} & \ge & m-\sum_{1\le j\le i}a_j\frac{n_{i-j}}{n_{i-1}} &\\
%  & = & \sum_{j\ge1}ja_jx_0^{j-1}-\sum_{1\le j\le i}a_j\frac{n_{i-j}}{n_{i-1}} &\\
%  & = & \sum_{1\le j\le i}ja_jx_0^{j-1}-\sum_{1\le j\le i}a_j\frac{n_{i-j}}{n_{i-1}}+\sum_{j>i}ja_jx_0^{j-1} &\\
%  & = & \sum_{1\le j\le i}ja_jx_0^{j-1}-\sum_{1\le j\le i}a_j\frac{n_{i-j}}{n_{i-1}}+\sum_{j>i}ja_jx_0^{j-1} &\\
%  & \ge & m-\sum_{1\le j\le i}\frac{x}{y}a_jx_0^{j-1} & \text{ By induction }\\
%  & \ge & m-\sum_{j\ge1}a_jx_0^{j-1} & \\
%  & = & x_0^{-1} & \text{  }.
% \end{array}$$

The power series method used in previous papers~\cite{BellGoh:2007,Blanchet:2013,Rampersad:2011}
corresponds to the special case such that $s_f=|f|$ for every forbidden factor.
Our condition is that $P(x)=0$ for some $x>0$ whereas the condition in these papers is that
every coefficient of the series expansion of $\frac1{P(x)}$ is positive.
The two conditions are actually equivalent.
The result in~\cite{Piotkovskii:1993} concerns series of the form $S(x)=1+a_1x+a_2x^2+a_3x^3+\ldots$ with real coefficients such that $a_1<0$
and $a_i\ge0$ for every $i\ge2$. It states that every coefficient of the series $1/S(x)=b_0+b_1x+b_2x^2+b_3x^3+\ldots$
is positive if and only if $S(x)$ has a positive real root $x_0$. Moreover, we have $b_i\ge x_0^{-i}$ for every $i\ge0$.

The entropy compression method as developped by Gon\c{c}alves, Montassier, and Pinlou~\cite{GMP:2015}
uses a condition equivalent to $P(x)=0$. The benefit of the present method is that we get an exponential
lower bound on the factor complexity. It is not clear whether it is possible to get such a lower bound
when using entropy compression for graph coloring, since words have a simpler structure than graphs.

\section{Applying the method}\label{sec:10}
%%%%%%%%%%%%%%%%%%%%%

In this section, we show that some doubled patterns on 4 and 5 variables are 3-avoidable.
Given a pattern $p$, every occurrence $f$ of $p$ is a forbidden factor.
With an abuse of notation, we denote by $|A|$ the length of the image of the variable $A$ of $p$
in the occurrence $f$. This notation is used to define the length $s_f$.

Let us first consider doubled patterns with 4 variables.
We begin with patterns of length 9, so that one variable, say $A$, appears 3 times.
We set $s_f=|f|$. Using the obvious upper bound on the number of pattern occurrences,
we obtain
$$\begin{array}{lll}
P(x) & = & 1-3x+\sum_{a,b,c,d\ge1}3^{a+b+c+d}x^{3a+2b+2c+2d}\\
 & = & 1-3x+\sum_{a,b,c,d\ge1}\paren{3x^3}^a\paren{3x^2}^b\paren{3x^2}^c\paren{3x^2}^d\\
 & = & 1-3x+\paren{\sum_{a\ge1}\paren{3x^3}^a}\paren{\sum_{b\ge1}\paren{3x^2}^b}\paren{\sum_{c\ge1}\paren{3x^2}^c}\paren{\sum_{d\ge1}\paren{3x^2}^d}\\
 & = & 1-3x+\paren{\frac{1}{1-3x^3}-1}\paren{\frac{1}{1-3x^2}-1}\paren{\frac{1}{1-3x^2}-1}\paren{\frac{1}{1-3x^2}-1}\\
 & = & 1-3x+\paren{\frac{1}{1-3x^3}-1}\paren{\frac{1}{1-3x^2}-1}^3\\
 & = & \frac{1-3x-9x^2+24x^3+36x^4-54x^5-108x^6+243x^8+162x^9-243x^{10}}{(1-3x^3)(1-3x^2)^3}.
\end{array}$$

%$P(x)=1-3x+\sum_{a,b,c,d\ge1}3^{a+b+c+d}x^{3a+2b+2c+2d}=1-3x+\paren{\frac1{1-3x^3}-1}\paren{\frac1{1-3x^2}-1}^3$.
Then $P(x)$ admits $x_0=0.3400\dots$ as its smallest positive real root.
So, every doubled pattern $p$ with 4 variables and length 9 is 3-avoidable and
there exist at least $x_0^{-n}>2.941^n$ ternary words avoiding $p$.
Notice that for patterns with 4 variables and length at least 10, every term of $\sum_{a,b,c,d\ge1}3^{a+b+c+d}x^{3a+2b+2c+2d}$
in $P(x)$ gets multiplied by some positive power of $x$. Since $0<x<1$, every term is now smaller than in
the previous case. So $P(x)$ admits a smallest positive real root that is smaller than $0.3400\dots$
Thus, these patterns are also 3-avoidable.

Now, we consider patterns with length 8, so that every variable appears exactly twice.
If such a pattern has $ABCD$ as a prefix, then we set $s_f=\frac{|f|}2=|A|+|B|+|C|+|D|$.
So we obtain $P(x)=1-3x+\sum_{a,b,c,d\ge1}x^{a+b+c+d}=1-3x+\paren{\frac1{1-x}-1}^4$.
Then $P(x)$ admits $0.3819\dots$ as its smallest positive real root, so that this pattern is 3-avoidable.

Among the remaining patterns, we rule out patterns containing an occurrence of a doubled pattern with at most 3 variables.
Also, if one pattern is the reverse of another, then they have the same avoidability index and we consider only one of the two.
Thus, there remain the following patterns:
$ABACBDCD$, $ABACDBDC$, $ABACDCBD$, $ABCADBDC$, $ABCADCBD$, $ABCADCDB$, and $ABCBDADC$.\\

Now we consider doubled patterns with 5 variables.
Similarly, we rule out every pattern of length at least 11 with the method
by setting $s_f=|f|$.
Then we check that $P(x)=1-3x+\sum_{a,b,c,d,e\ge1}3^{a+b+c+d+e}x^{3a+2b+2c+2d+2e}=1-3x+\paren{\frac1{1-3x^3}-1}\paren{\frac1{1-3x^2}-1}^4$
has a positive real root.

We also rule out every pattern of length 10 having $ABC$ as a prefix.
We set $s_f=|f|-|ABC|=|A|+|B|+|C|+2|D|+2|E|$.
Then we check that $P(x)=1-3x+\sum_{a,b,c,d,e\ge1}3^{d+e}x^{a+b+c+2d+2e}=1-3x+\paren{\frac1{1-x}-1}^3\paren{\frac1{1-3x^2}-1}^2$
has a positive real root.

Again, we rule out patterns containing an occurrence of a doubled pattern with at most 4 variables
and patterns whose reversed pattern is already considered.
Thus, there remain the following patterns: $ABACBDCEDE$, $ABACDBCEDE$, and $ABACDBDECE$.

%%%%%%%%%%%%%%%%%%%%%
\section{Sporadic doubled patterns}\label{sec:10}
%%%%%%%%%%%%%%%%%%%%%
In this section, we consider the 10 doubled patterns on 4 and 5 variables whose 3-avoidability
has not been obtained in the previous section.
%We use the method described in~\cite{Ochem:2004} to show that they are 2-avoidable.

We define the \emph{avoidability exponent} $AE(p)$ of a pattern $p$ as the largest real $x$ such
that every $x$-free word avoids $p$. This notion is not pertinent e.g. for the pattern $ABWBAXACYCAZBC$ studied
by Baker, McNulty, and Taylor~\cite{BNT89}, since for every $\epsilon>0$, there exists a
$(1+\epsilon)$-free word containing an occurrence of that pattern.
However, $AE(p)>1$ for every doubled pattern. To see that, consider a factor $A\dots A$ of $p$.
If an $x$-free word contains an occurrence of $p$, then the image of this factor is a repetition such that the image of $A$
cannot be too large compared to the images of the variables occurring between the $A$s in $p$.
We have similar constraints for every variable and this set of constraints becomes unsatisfiable as $x$ decreases towards 1.
We present one way of obtaining the avoidability exponent for a doubled pattern $p$ of length $2v(p)$.
We construct the $v(p)\times v(p)$ matrix $M$ such that $M_{i,j}$ is the number of occurrences of the variable $X_j$ between
the two occurrences of the variable $X_i$.
We compute the largest eigenvalue $\beta$ of $M$ and then we have $AE(p)=1+\frac1{\beta+1}$.
For example if $p=ABACDCBD$, then we get $M=\cro{\begin{smallmatrix} 0 & 1 & 0 & 0\\ 1 & 0 & 0 & 1\\ 0 & 2 & 0 & 1\\ 0 & 1 & 1 & 0\end{smallmatrix}}$,
$\beta=1.9403\dots$, and $AE(p)=1+\frac1{\beta+1}=1.3400\dots$.
The avoidability exponents of the 10 patterns considered in this section range from $AE(ABCADBDC)=1.292893219$ to
$AE(ABACBDCD)=1.381966011$.
For each pattern $p$ among the 10, we give a uniform morphism $m:\Sigma_5^*\to\Sigma_2^*$ such that for every $\paren{\frac54^+}$-free word $w\in\Sigma_5^*$,
we have that $m(w)$ avoids $p$. The proof that $p$ is avoided follows the method in~\cite{Ochem:2004}.
Since there exist exponentially many $\paren{\frac54^+}$-free words
over $\Sigma_5$~\cite{KR:2011}, there exist exponentially many binary words avoiding $p$.

\begin{itemize}
 \item 
$AE(ABACBDCD)=1.381966011$, 17-uniform morphism
$$
\begin{array}{c}
0\mapsto 00000111101010110\\
1\mapsto 00000110100100110\\
2\mapsto 00000011100110111\\
3\mapsto 00000011010101111\\
4\mapsto 00000011001001011\\
\end{array}
$$

 \item 
$AE(ABACDBDC)=1.333333333$, 33-uniform morphism
$$
\begin{array}{c}
0\mapsto 000000101101000111111011001010111\\
1\mapsto 000000100110100001111101001010111\\
2\mapsto 000000010110100001111111010010111\\
3\mapsto 000000010011010100011111010010111\\
4\mapsto 000000010011001000001111010010111\\
\end{array}
$$

 \item 
$AE(ABACDCBD)=1.340090632$, 28-uniform morphism
$$
\begin{array}{c}
0\mapsto 0000101010001110010000111111\\
1\mapsto 0000001111010001101001111111\\
2\mapsto 0000001101000011110100100111\\
3\mapsto 0000001011110000110100111111\\
4\mapsto 0000001010111100100001111111\\
\end{array}
$$

 \item 
$AE(ABCADBDC)=1.292893219$, 21-uniform morphism
$$
\begin{array}{c}
0\mapsto 000011101101011111010\\
1\mapsto 000010110100100111101\\
2\mapsto 000001101110100101111\\
3\mapsto 000001101011001111111\\
4\mapsto 000000110111010111111\\
\end{array}
$$

 \item 
$AE(ABCADCBD)=1.295597743$, 22-uniform morphism
$$
\begin{array}{c}
0\mapsto 0000011011010100011111\\
1\mapsto 0000011010101001001111\\
2\mapsto 0000001101100100111111\\
3\mapsto 0000001010110000111111\\
4\mapsto 0000000110101001110111\\
\end{array}
$$

 \item 
$AE(ABCADCDB)=1.327621756$, 26-uniform morphism
$$
\begin{array}{c}
0\mapsto 00000011110010101011000111\\
1\mapsto 00000011010111111001011011\\
2\mapsto 00000010011111101001110111\\
3\mapsto 00000001001111110001010111\\
4\mapsto 00000001000111111001010111\\
\end{array}
$$

 \item 
$AE(ABCBDADC)=1.302775638$, 33-uniform morphism
$$
\begin{array}{c}
0\mapsto 000000101111110011000110011111101\\
1\mapsto 000000101111001000001100111111101\\
2\mapsto 000000011011111001100000100111101\\
3\mapsto 000000011010101011000001001111101\\
4\mapsto 000000010111110010101010011111011\\
\end{array}
$$

 \item 
$AE(ABACBDCEDE)=1.366025404$, 15-uniform morphism
$$
\begin{array}{c}
0\mapsto 001011011110000\\
1\mapsto 001010100111111\\
2\mapsto 000110010011000\\
3\mapsto 000011111111100\\
4\mapsto 000011010101110\\
\end{array}
$$

 \item 
$AE(ABACDBCEDE)=1.302775638$, 18-uniform morphism
$$
\begin{array}{c}
0\mapsto 000010110100100111\\
1\mapsto 000010100111111111\\
2\mapsto 000000110110011111\\
3\mapsto 000000101010101111\\
4\mapsto 000000000111100111\\
\end{array}
$$

 \item 
$AE(ABACDBDECE)=1.320416579$, 22-uniform morphism
$$
\begin{array}{c}
0\mapsto 0000001111110001011011\\
1\mapsto 0000001111100100110101\\
2\mapsto 0000001111100001101101\\
3\mapsto 0000001111001001011100\\
4\mapsto 0000001111000010101100\\
\end{array}
$$
\end{itemize}

%%%%%%%%%%%%%%%%%%%%%
\section{Simultaneous avoidance of doubled patterns}\label{sec:sim}
%%%%%%%%%%%%%%%%%%%%%

Bell and Goh~\cite{BellGoh:2007} have also considered the avoidance of multiple patterns simultaneously
and ask (question 3) whether there exist an infinite word over a finite alphabet that avoids every doubled pattern.
We give a negative answer.

% Let us assume that such a word $w$ exists.
% Without loss of generality, we assume that $w$ is uniformly reccurrent, that is, for every finite factor $q$ of $w$,
% there exists an integer $\ell_q$ such that every factor of $w$ of length $\ell_q$ contains $q$.
% Indeed, if $w$ is not uniformly reccurrent then $w$ contains a factor $q$
% such that $w$ contains arbitrarily large factors that do not contain $q$.
% By compacity, there exists an infinite word $w'$ such that every factor of $w'$ is a factor of $w$
% and $q$ is not a factor of $w'$.
% Since $w$ is reccurrent, we can define $\ell=2\max\acc{\ell_a}$ where the maximum runs over the letters $a$ of $w$.
% By definition, the prefix $u$ of length $\ell$ of $w$ contains every letter at least twice.
% So $u$ is an occurrence of a doubled pattern such that every variable has a distinct image of length~1.
% This contradiction shows that no infinite word over a finite alphabet avoids every doubled pattern.
% We can prove more.

A word $w$ is \emph{$n$-splitted} if $|w|\equiv 0\pmod{n}$ and every factor $w_i$ such that
$w=w_1w_2\ldots w_n$ and $|w_i|=\tfrac{|w|}n$ for $1\le i\le n$ contains every letter in $w$.
An $n$-splitted pattern is defined similarly.
Let us prove by induction on $k$ that every word $w\in\Sigma_k^{n^k}$ contains an $n$-splitted factor.
The assertion is true for $k=1$.
Now, if the word $w\in\Sigma_k^{n^k}$ is not itself $n$-splitted, then by definition it must contain a factor $w_i$
that does not contain every letter of $w$. So we have $w_i\in\Sigma_{k-1}^{n^{k-1}}$.
By induction, $w_i$ contains an $n$-splitted factor, and so does $w$.

This implies that for every fixed $n$, every infinite word over a finite alphabet contains
$n$-splitted factors. Moreover, an $n$-splitted word is an occurrence of an $n$-splitted pattern
such that every variable has a distinct image of length~1.
So, for every fixed $n$, the set of all $n$-splitted patterns is not avoidable by
an infinite word over a finite alphabet.

Notice that if $n\ge 2$, then an $n$-splitted word (resp. pattern) contains a $2$-splitted word (resp. pattern)
and a $2$-splitted word (resp. pattern) is doubled.

%%%%%%%%%%%%%%%%%%%%%
\section{Conclusion}\label{sec:con}
%%%%%%%%%%%%%%%%%%%%%

Our results answer settles the first of two questions of our previous paper~\cite{OchemPinlou}.
The second question is whether there exists a finite $k$ such that every doubled pattern
with at least $k$ variables is 2-avoidable. As already noticed~\cite{OchemPinlou},
such a $k$ is at least 5 since, e.g., $ABCCBADD$ is not 2-avoidable.

%%%%%%%%%%%%%%%%%%%%%%%
\section*{Acknowledgments}
%%%%%%%%%%%%%%%%%%%%%%%

I am grateful to Narad Rampersad for comments on a draft of the paper,
to Vladimir Dotsenko for pointing out the result in~\cite{Piotkovskii:1993},
and to Andrei Romashchenko for translating this paper.

\end{document}